\documentclass[12pt]{article}
\usepackage[latin1]{inputenc} 
\usepackage{amssymb}
\usepackage{amsfonts} 
\usepackage{amsthm} 
\usepackage{amsmath} 
\usepackage{hyperref}
\usepackage{color}
\usepackage{bbold}
\usepackage{young}

\def\nn{\nonumber}
\def\ic{\mathrm{i}}

\def \bc {\begin{center}}
\def \ec {\end{center}}
\def \bi {\begin{itemize}}
\def \ei {\end{itemize}}

\def \ba {\begin{array}}
\def \ea {\end{array}}

\def \bea {\begin{eqnarray}}
\def \eea {\end{eqnarray}}

\def \be {\begin{equation}}
\def \ee {\end{equation}}

\def \lb {\left[}
\def \rb {\right]}

\newcommand{\ra}{\rangle}

\def \um {\frac{1}{2}}

\def\tr {\mathrm{tr}}

\def\cL {{\cal L}}

\def\rmu {\mathrm{U}}
\def\rmuu {\mathrm{u}}

\def\rmsu {\mathrm{SU}}

\newtheorem{thm}{Theorem}[section]

\newtheorem{prop}[thm]{Proposition}
\newtheorem{defn}[thm]{Definition}
\theoremstyle{remark}
\newtheorem{rem}[thm]{Remark}

\begin{document}

\begin{center}
{\Large {\bf Massive conformal particles  with non-Abelian charges from free U(2N,2N)-twistor dynamics: quantization and coherent states}}
\end{center}
\bigskip

\centerline{{\sc Manuel Calixto}\footnote{calixto@ugr.es}}

\bigskip

\bc {\it  Department of Applied Mathematics and  Institute ``Carlos I'' of Theoretical and Computational Physics,
Faculty of Sciences, University of Granada, Fuentenueva s/n,  18071 Granada, Spain}
\ec

\bigskip
\begin{center}
{\bf Abstract}
\end{center}
\small 
\begin{list}{ }{\setlength{\leftmargin}{3pc}\setlength{\rightmargin}{3pc}}
\item 
The inclusion of non-Abelian $\rmu(N)$ internal charges (other than the electric charge) into Twistor Theory is accomplished through the concept of 
``colored twistors'' (ctwistors for short) transforming under the colored conformal symmetry $\rmu(2N,2N)$. In particular, we are interested in $2N$-ctwistors describing 
colored spinless conformal massive particles with phase space $\rmu(2N,2N)/\rmu(2N)\times\rmu(2N)$. Penrose formulas for incidence relations are generalized to $N>1$. 
We propose $\rmu(2N)$-gauge invariant Lagrangians for 
$2N$-ctwistors  and we quantize them through a bosonic representation, interpreting quantum states as particle-hole excitations above the ground state. 
The connection between the corresponding Hilbert (Fock-like with constraints) space and the carrier space of a discrete 
series representation of $\rmu(2N,2N)$ is established through a coherent state (holomorphic) representation. 
\end{list}
\normalsize 




\section{Introduction\label{intro}}

A classical description of particles with non-Abelian charges was given long ago by Wong \cite{Wong} 
in terms of equations of motion, and the Lagrangian and Hamiltonian descriptions of
the corresponding dynamics was formulated in \cite{Barducci} and \cite{Bal}. Here we are interested in the 
formulation of a theory of particle (electromagnetic, weak and strong) interactions in twistor terms. 
The Twistor Program was  introduced by R. Penrose and coworkers 
\cite{Penrose,Penrose2,Penrose3,Penrose4,Penrose5} in the 1960's  as an approach to the unification of quantum theory with gravity. 
Penrose, Perj\'es \cite{Perjes,Perjes2,Perjes3,Perjes4}, and Hughston 
\cite{Hughston} made some attempts to formulate models of 
massive spinning particles with internal symmetries in Minkowski space in terms of $M$-twistors, proposing the identification of 
the SU(2) and SU(3) symmetries appearing in the $M=2$ and $M=3$ twistor models with the symmetry for leptons and hadrons, respectively. 
In particular, Hughston \cite{Hughston} studied the  three-twistor model for low-lying barions and mesons carrying electric charge, 
hypercharge, barion number and isospin. However, there is an absence of color (only flavor) degrees of freedom. Also, the 
inherent chirality of twistor space seemed to be a handicap, until 2004 Witten's paper \cite{Witten} on 
twistorial representations of scattering amplitudes showed how to overcome this issue when string theory is introduced into twistor space. This 
entails a spur to pursue a reworking of particle interactions in twistor language and it is the subject of much recent activity 
(see e.g. \cite{Hodges} and references therein).

The Lagrangian mechanics of a massive spinning particle in Minkowski space, formulated in terms of 2-twistors, 
has been revisited  in \cite{PLB595,IJGMMP,PRD73,PLB733,JHEP} (see \cite{spinning} for other formulations in terms of 
nonlinear sigma models on U(2,2)). The sixteen real coordinates of two-twistor space are transformed 
into an enlarged relativistic phase space-time framework, containing  
the standard relativistic phase space of coordinates supplemented by a six-parameter spin  and a two-parameter electric charge phase space, 
with constraints. Only the two-twistor case is adopted, without mention to the $M(\geq 3)$-twistor descriptions. 
Actually, it is shown in \cite{Townsend}  that only the two-twistor formulation can successfully describe a massive particle in 
Minkowski space. Moreover, it is proved in \cite{nogotwistor} that the $M$-twistor expression of a particle's four-momentum vector reduces 
to the two-twistor expression for a massive particle or the one-twistor expression for a massless particle. Therefore, they conclude that 
the genuine $M$-twistor description of a massive particle in four-dimensional Minkowski space fails for the case $M\geq  3$. This is 
a kind of ``no-go'' theorem that prevents the inclusion of internal charges (other than the electric charge) into the model, as originally 
envisaged by \cite{Penrose,Penrose2,Penrose3,Penrose4,Penrose5,Perjes,Perjes2,Perjes3,Perjes4,Hughston}.

In this article we pursue a way out to this no-go theorem by replacing standard twistors by \emph{colored twistors} (``ctwistors'' for short), 
which transform under the \emph{colored conformal group} $\rmu(2N,2N)$, with $N$ the number of colors of our particle. Although strong interactions 
suggests $N=3$, we shall leave $N$ arbitrary all along the paper for the sake of generality, comparing with the standard $N=1$ case. 
Colored twistors have enough room to accommodate non-Abelian internal degrees of freedom. In particular, we are interested in $2N$-ctwistors describing 
colored spinless conformal massive particles with $8N^2$-dimensional (``complex colored Minkowski'') phase space $\rmu(2N,2N)/\rmu(2N)\times\rmu(2N)$, which reduces 
to the forward tube domain of the complex Minkowski space for $N=1$. In this article we shall analyze the structure of the 
colored conformal symmetry $\rmu(2N,2N)$ (its discrete series representations) and the quantization of colored conformal massive particles 
through a bosonic representation of colored twistors. 

The organization of the paper is the following. In Section \ref{colorconf} we describe the colored conformal symmetry (Lie algebra generators and coordinate systems) 
and its discrete series representations, which provides the carrier Hilbert space of our colored conformal massive particle. In Section \ref{twistnlsm} we formulate nonlinear 
sigma model Lagrangians  on $\rmu(2N,2N)/\rmu(2N)\times\rmu(2N)$ (either as a phase or a configuration space) for colored conformal massive particles in terms of colored twistors. 
They are $\rmu(2N)$-gauge invariant and their quantization is accomplished in Section \ref{bosonsec} in a Fock space (bosonic) picture with constraints. The basic quantum states of the 
corresponding Hilbert space are constructed by repeated application of particle-hole (``exciton'') ladder operators on the ground state.  The connection between this bosonic representation and the 
holomorphic representation offered at the end of Section \ref{colorconf} is achieved through a coherent state representation of the corresponding quantum states. 
In Section \ref{fieldtsec} we briefly extend these ideas to the many-particle case, by formulating  field theory Lagrangian densities. Finally, Section \ref{conclusec} is devoted to 
conclusions and outlook.

\section{Colored conformal symmetry and discrete series representation\label{colorconf}}
In this Section we describe the underlying symmetry $\rmu(2N,2N)$, Lie algebra structure, irreducible representations and infinitesimal generators.

\subsection{Colored conformal group and complex Minkowski space}

Let us start by discussing the group theoretical backdrop, fixing notation and reminding some standard definitions. 
The unitary group $\mathrm{U}(M)$ is a subgroup of the general linear group $\mathrm{GL}(M,\mathbb C)$ fulfilling:
\be \mathrm{U}(M)=\left\{U\in \mathrm{GL}(M,\mathbb C):   U^\dag  U= U U^\dag=\mathbb{1}_M\right\},\label{pseudou} \ee
where $\dag$ means conjugate transpose and $\mathbb{1}_M$ is the ${M\times M}$ identity matrix (we use the fundamental or defining representation for the sake of convenience). 
The $\mathrm{u}(M)$ Lie algebra basis is then given by the $M^2$ hermitian matrices 
[we use the physicist convention for the exponential map $U=\exp(\ic \lambda)$, with $\lambda$ hermitian and $\ic$ the imaginary unit]:
\bea
(\lambda^{ij})_{kl}&=&(\delta_{ik}\delta_{jl}+\delta_{il}\delta_{jk})/\sqrt{2},\; j>i=1,\dots,M,\nn\\
(\lambda^{ji})_{kl}&=&-\ic(\delta_{ik}\delta_{jl}-\delta_{il}\delta_{jk})/\sqrt{2},\; j>i=1,\dots,M,\nn\\
\lambda^i &=& \frac{1}{\sqrt{i(i+1)}}\mathrm{diag}(1,\stackrel{i}{\dots},1,-i,0,\stackrel{M-i-1}{\dots},0),\; i=1,\dots,M-1. \label{PauliN}
\eea
[which generate $\mathrm{su}(M)$] plus the ${M\times M}$ identity matrix $\mathbb{1}_M=\sqrt{M}\lambda^0$ (the `trace' or linear Casimir). 
The matrices $\lambda$ constitute 
a generalization of the usual Pauli matrices $\sigma$ from $\mathrm{SU}(2)$ to $\mathrm{SU}(M)$ [more precisely, $\sigma^1=\sqrt{2}\lambda^{12}, 
\sigma^2=\sqrt{2}\lambda^{21}$ and $\sigma^3=\sqrt{2}\lambda^1$] plus the identity $\sigma^0=\sqrt{2}\lambda^0$. 
We shall denote all the $\mathrm{U}(M)$ Lie algebra generators collectively 
by $\lambda^a, a=0,\dots,M^2-1$, so that the Killing form is simply 
$\tr(\lambda^a\lambda^b)=\delta^{ab}$, which is a generalization of the usual relation $\tr(\sigma^\mu\sigma^\nu)=2\delta^{\mu\nu}$ for $\rmu(2)$ Pauli matrices.  

This Lie-algebra structure is straightforwardly translated to the non-compact counterpart (for even $M$) 
\be
\rmu(M/2,M/2)=\left\{ \tilde{U}\in {\rm Mat}_{M\times M}(\mathbb C):  \tilde{U}^\dag \Gamma \tilde{U}=\Gamma\right\},\label{suMM} \ee
of pseudo-unitary matrices $\tilde{U}$ leaving invariant the $M\times M$ Hermitian form $\Gamma$ of signature $(1,\stackrel{M/2}{\dots},1,-1,\stackrel{M/2}{\dots},-1)$. 
The Lie algebra is made of pseudo-hermitian matrices $\tilde{\lambda}$ fulfilling $\tilde{\lambda}^\dag=\Gamma\tilde{\lambda}\Gamma$; 
actually, if $\lambda$ is a (hermitian) generator of $\rmu(M)$, then $\tilde\lambda=\Gamma\lambda$ is a (pseudo-hermitian) generator of $\rmu(M/2,M/2)$. 

For the colored conformal symmetry $\rmu(2N,2N)$ we will rather prefer sometimes a different Lie-algebra basis adapted to the usual fundamental matrix realization of the 
$\rmsu(2,2)$ conformal generators in terms of fifteen $4\times 4$ matrices $D, M^{\mu\nu}, P^\mu$ and $K^\mu$ (dilation, Lorentz, translation and acceleration, respectively) 
of the form
\be  \ba{cc}  D=\frac{1}{2}\left(\ba{cc} -\sigma^0 & 0\\ 0 &\sigma^0 \ea\right), &
M^{\mu\nu}=\frac{1}{4}\left(\ba{cc} \sigma^\mu\check{\sigma}^\nu-\sigma^\nu\check{\sigma}^\mu & 0\\
0&\check{\sigma}^\mu\sigma^\nu-\check{\sigma}^\nu\sigma^\mu\ea\right),  \\ 
 P^\mu=\left(\ba{cc} 0& \sigma^\mu \\ 0
&0\ea\right), & K^\mu=\left(\ba{cc} 0& 0 \\ 
 \check{\sigma}^\mu &0\ea\right).\ea  \label{confalgamma}\ee
where $\check{\sigma}^\mu\equiv \sigma_\mu=\eta_{\mu\nu}\sigma^\nu$ [we are using the
convention $\eta={\rm diag}(1,-1,-1,-1)$ for the Minkowski metric] denote parity-reversed Pauli matrices. Denoting the sixteen $\rmu(2,2)$ generators 
$\{\mathbb{1}_4,D, M^{\mu\nu}, P^\mu, K^\mu\}$ collectively 
by $\{T^\alpha, \alpha=0,\dots, 15\}$, a natural basis for the colored conformal symmetry $\rmu(2N,2N)$ is the direct product
\be
T^{\alpha a}=T^\alpha\otimes \lambda^a,\; \alpha=0,\dots, 15,\; a=0,\dots,N^2-1\label{colconfalg}
\ee
of  space-time $\rmu(2,2)$ symmetry generators  $T^\alpha$ times internal $\rmu(N)$ symmetry generators $\lambda^a$. 

Let us define the colored twistor space $\mathbb{C}^{4N}$ as the basic representation space of the colored conformal group $\rmu(2N,2N)$. 
Lines, planes, etc, in $\mathbb{C}^{4N}$ lead to the notion of (ctwistor) flag manifolds, which can be regarded as homogeneous spaces of $\rmu(2N,2N)$. 
There is a one-to-one correspondence between orbits 
\be\mathbb{O}_{\tilde\lambda}=\{\Lambda=\mathrm{Ad}_{U}(\tilde\lambda)=U\tilde\lambda U^\dag, \, U\in \mathrm{U}(2N,2N)\}, \ee
of the adjoint representation of $\mathrm{U}(2N,2N)$ on a given 
Lie-algebra $\rmuu(2N,2N)$ generator $\tilde\lambda$ (usually in the Cartan subalgebra of diagonal matrices), and  cosets $\mathrm{U}(2N,2N)/H_{\tilde{\lambda}}$, 
with  $H_{\tilde{\lambda}}$  the isotropy group (stabilizer) of $\tilde{\lambda}$. In particular, we are interested in the Grassmann manifold 
of $2N$ planes in $\mathbb{C}^{4N}$, which is the colored conformal orbit 
\be \mathbb{D}_{4N^2}=\mathrm{U}(2N,2N)/[\rmu(2N)\times\rmu(2N)]\ee
of the adjoint action of $\mathrm{U}(2N,2N)$ on the dilation generator 
\be \tilde\lambda=\tilde D=D\otimes \lambda_0\propto(1,\stackrel{2N}{\dots},1,-1,\stackrel{2N}{\dots},-1)\ee
with isotropy group $H_{\tilde{D}}=\rmu(2N)\times\rmu(2N)$ the maximal compact subgroup of $\mathrm{U}(2N,2N)$. 
For example, for $N=1$, The conformal Cartan-Bergman domain $\mathbb{D}_{4}$ is eight-dimensional and corresponds to the phase space of a 
spinless positive mass conformal particle \cite{Todorov,Coquereaux}. A parametrization of the complex manifold $\mathbb{D}_{4N^2}$ 
can be obtained as follows. Any group element $U\in \mathrm{U}(2N,2N)$ (in a given patch, containing the identity element) admits the Iwasawa
decomposition (in block matrix form)
\be U=\left(\ba{cc} A& C
\\ B &D\ea\right)=\left(\ba{cc} \Delta_1& Z^\dag\Delta_2
\\ Z\Delta_1 &\Delta_2\ea\right)\left(\ba{cc} V_1& 0
\\ 0 &V_2\ea\right),\; \label{Iwasawa}\ee
with
\be \ba{lll} Z=BA^{-1}, & \Delta_1=(AA^\dag)^{\frac{1}{2}}=(\mathbb{1}_{2N}-Z^\dag Z)^{-\frac{1}{2}}, & V_1=(AA^\dag)^{-\frac{1}{2}}A, \\ 
 Z^\dag=CD^{-1}, & \Delta_2=(DD^\dag)^{\frac{1}{2}}=(\mathbb{1}_{2N}-Z Z^\dag)^{-\frac{1}{2}}, & 
V_2=(DD^\dag)^{-\frac{1}{2}}D. \ea \label{cosetcoord} \ee
Note that  $V_1, V_2\in \rmu(2N)$ are unitary matrices and each $2N\times 2N$ complex matrix  $Z$ defines an equivalence class representative of the quotient 
$\mathrm{U}(2N,2N)/\rmu(2N)^2$ {[See later on Remark \ref{incidence}, which relates this ``gauge fixing'' $Z=BA^{-1}$ to the solution 
of ``colored'' Penrose incidence equations]}. Therefore, the colored conformal Cartan domain can be defined through the positivity condition
\be
\mathbb{D}_{4N^2}=\{Z\in\mathrm{Mat}_{2N\times 2N}(\mathbb{C}): \mathbb{1}_{2N}-Z^\dag Z>0\}.
\ee
The Shilov boundary of $\mathbb{D}_{4N^2}$ (that is, those points fulfilling $Z^\dag Z=\mathbb{1}_{2N}$) is then the 
$4N^2$-dimensional ``compactified colored Minkowski space'' $\rmu(2N)$. Actually, for $N=1$, the group manifold of $\rmu(2)$ is the compactification 
$(\mathbb{S}^3\times\mathbb{S}^1)/\mathbb{Z}_2$ of the four-dimensional Minkowski space $\mathbb{R}^4$.

Let us introduce a suitable set of $4N^2$ complex (colored Minkowski) coordinates $z_{\mu a}$ on $\mathbb{D}_{4N^2}$  using the $\rmu(2N)$ 
Lie algebra matrices $\sigma^\mu\otimes \lambda^a$ as
\be Z=z_{\mu a}\sigma^\mu\otimes \lambda^a, \; z_{\mu a}=\frac{1}{2}\tr(Z\sigma^\mu\otimes \lambda^a), \; \mu=0,\dots,3, \; a=0,\dots,N^2-1.\label{ccc}
\ee
The particular case $z_\mu=z_{\mu 0}$ provides a coordinate system of the eight-dimensional Cartan domain $\mathbb{D}_{4}$, which can be mapped onto the 
forward tube domain 
\be
\mathbb{T}_4=\{W=X+\ic Y\in\mathrm{Mat}_{2\times 2}(\mathbb{C}):Y>0\}
\ee
of the complex Minkowski (phase) space $\mathbb{C}^4$, with $X=x_\mu\sigma^\mu$ (four-position) and $Y=y_\mu\sigma^\mu$ (four-momenta) hermitian matrices. 
{To be more presise,  position $x_\mu$ and momenta $y_\nu$ are conjugated but  not ``canonically congugated''. However, we can define  a proper 
canonically conjugated four-momentum as $p_\nu=y_\nu/y^2$, which gives the desired (canonical) Poisson bracket $\{x_\mu,p_\nu\}=\eta_{\mu\nu}$, when we look at 
$\mathbb{T}_4$ as a K\"ahler manifold with a closed (symplectic) two form (see \cite{spinning} for more details). From this point of view, the positivity condition  
$Y>0\Leftrightarrow y_0>\|\vec{y}\|$ is related to the positive energy condition $p_0>\|\vec{p}\|$}. 
The  mapping from $\mathbb{D}_4$ onto $\mathbb{T}_4$ 
is the four-dimensional analogue of the usual M\"obius map $z\to(z+\ic)/(\ic z+1)$ from the unit disk onto the upper half-plane in two dimensions. 
We can extend this map (also denoted by Cayley transform) to 
$\mathbb{D}_{4N^2}$  and $\mathbb{T}_{4N^2}$ by defining the ``colored Cayley transform''  and its inverse as
\bea 
Z&\to& W=\ic (\mathbb{1}_{2N}-Z)(\mathbb{1}_{2N}+Z)^{-1}, \nn\\
W&\to& Z= (\mathbb{1}_{2N}-\ic W)^{-1}(\mathbb{1}_{2N}+\ic W).
\eea
For convenience, we shall rather prefer the Cartan domain $\mathbb{D}_{4N^2}$ to the forward tube domain $\mathbb{T}_{4N^2}$ picture, for the phase space of our 
colored massive conformal particle, although we perhaps have more physical intuition in $\mathbb{T}_4$ than in $\mathbb{D}_4$ for $N=1$. 
{Note that in the phase space $\mathbb{T}_{4N^2}\ni W=X+\ic Y$ of our colored conformal spinless massive particle,  we have $4N^2$ ``colored position''  
$x_{\mu a}=\tr(X\sigma^\mu\otimes \lambda^a)/2$ and $4N^2$ ``colored momentum''  
$y_{\mu a}=\tr(Y\sigma^\mu\otimes \lambda^a)/2$ coordinates. Pure spacetime coordinates correspond to $x_\mu=x_{\mu 0}$, which transform under the 
pure conformal transformations generated by $T^{\alpha 0}$ in \eqref{colconfalg}. The remainder generators mix spacetime and internal degrees of freedom, according 
to a  ``colored M\"obius'' transformation of phase-space coordinates $z_{\mu a}$ under general colored conformal transformations $U$ [see later on eq. \eqref{Ztransf}]. 
Note that this colored M\"obius transformation is nonlinear, unlike other approaches like \cite{Barducci,Bal} (also non conformal), which introduce coordinates 
$x_{\mu i}=(x_\mu,\theta_i), i=1,\dots,N$, to describe classical colored particles, where $\theta_i$ are Grasssmann variables 
transforming linearly under the fundamental representation of $\rmu(N)$, thus not allowing an interesting mixture between spacetime and internal degrees of 
freedom, like our approach does. The counting of phase-space degrees of freedom is also different in both approaches; indeed, we have much more room in our 
$8N^2$-dimensional phase space.
}

\subsection{\label{holopic}Discrete series representation of the colored conformal group} 

Let us now discuss the structure of the Hilbert space for our colored conformal massive particle as the carrier space of a unitary irreducible representation of 
the colored conformal group. Let us start by considering the Hilbert space $L^2(\rmu(2N,2N),d\mu)$ of square integrable complex functions $\psi(U)$ on 
$\rmu(2N,2N)\ni U$ with invariant scalar product 
\be
( \psi|\psi')=\int\overline{\psi(U)}\psi'(U)d\mu(U), \; d\mu(U)=\det(\mathbb{1}_{2N}-Z^\dag Z)^{-4N}|dZ| dv(V_1) d v(V_2),\label{spc}
\ee
given through the invariant Haar measure $d\mu(U)$ on $\rmu(2N,2N)$, which has been decomposed as the product of the invariant measure on  $\mathbb{D}_{4N^2}$ 
($|dZ|$ denotes the standard Lebesgue measure on $\mathbb{C}^{4N^2}$) and  the measure $dv$ on $\rmu(2N)\times \rmu(2N)$, 
according to the Iwasawa decomposition \eqref{Iwasawa}. The colored conformal group is unitarily represented in $L^2(\rmu(2N,2N),d\mu)$ by the left action
$[\mathcal{U}(U')\psi](U)=\psi(U'^{-1}U)$. However, this 
representation is highly reducible. As we want to restrict ourselves to the quotient $\mathbb{D}_{4N^2}=\rmu(2N,2N)/\rmu(2N)^2$, we can choose 
$\psi_\mathbb{n}(U)=\det(A)^{-\mathbb{n}}$ [for the decomposition \eqref{Iwasawa}] as the lowest-weight state, where $\mathbb{n}$ is an integer that will eventually 
label the representation (for $N=1$,  $\mathbb{n}$ corresponds to the \emph{scale or conformal dimension}, also related to the conformal invariant mass). 
The exponent of $\det(A)$ is chosen to be negative for irreducibility reasons (see below) and it is not a problem since $\det(A)\not=0$. 
Under $U'\in \rmu(2N)^2$ (the maximal compact subgroup), the lowest-weight state $\psi_\mathbb{n}$ remains invariant 
\be \psi_\mathbb{n}(U'^{-1}U)=\det(V_1^\dag A)^{-\mathbb{n}}=\det(V_1)^\mathbb{n}\psi_\mathbb{n}(U),\; U'=\left(\ba{cc} V_1& 0
\\ 0 &V_2\ea\right),\label{LWinv}\ee
up to an irrelevant phase $\det(V_1)^\mathbb{n}$ [note that, in this sense, there are other options like $\det(D)^{-\mathbb{n}}$ for the lowest-weight state]. 
Moreover, according to \eqref{cosetcoord}, the lowest-weight state modulus 
\[ |\psi_\mathbb{n}(U)|^2=\det(AA^\dag)^{-\mathbb{n}}=\det(\mathbb{1}_{2N}-Z^\dag Z)^{\mathbb{n}}
 \]
can also be just written in terms of coordinates $Z\in\mathbb{D}_{4N^2}$. Under a 
general colored conformal transformation $U'\in \rmu(2N,2N)$ the lowest-weight state transforms as
\be
[\mathcal{U}(U')\psi_\mathbb{n}](U)=\psi_\mathbb{n}(U'^{-1}U)=\det(A'^\dag A+B'^\dag B)^{-\mathbb{n}}=\det(A'^\dag+B'^\dag Z)^{-\mathbb{n}}\psi_\mathbb{n}(U),
\ee
where we have used that $U'^{-1}=\Gamma U'^\dag \Gamma=\left(\ba{cc} A'^\dag& -B'^\dag
\\ -C'^\dag &D'^\dag\ea\right)$ and the definition of $Z=BA^{-1 }$ in \eqref{cosetcoord}. The factor $\det(A'^\dag+B'^\dag Z)^{-\mathbb{n}}$ plays the role of a \emph{multiplier} and 
its expansion in holomorphic polynomial functions $\phi(Z)$ requires homogeneous polynomials of all order degrees (as long as $\mathbb{n}>0$), 
thus implying an infinite-dimensional representation of $\rmu(2N,2N)$, as required by unirreps of non-compact groups  (see \cite{EMSMTA} for a relation of this expansion with the MacMahon-Schwinger's Master Theorem). 
In other words, we have chosen $\det(A)^{-\mathbb{n}}$ [and not $\det(A)^{\mathbb{n}}$], with $\mathbb{n}>0$,  for unitarity reasons. 

The set of functions $\{ \psi^U_\mathbb{n}=\mathcal{U}(U)\psi_\mathbb{n}, U\in \rmu(2N,2N) \}$ in the orbit of the \emph{fiducial vector} 
$\psi_\mathbb{n}$ under $\rmu(2N,2N)$ is usually referred to as a \emph{coherent state system}. Note that $\psi^{U'}_\mathbb{n}(U)$ can also be written 
as
\be
\psi_\mathbb{n}(U'^{-1}U)=\overline{\psi_\mathbb{n}(U')}\mathfrak{K}_\mathbb{n}(Z'^\dag,Z)\psi_\mathbb{n}(U), \quad \mathfrak{K}_\mathbb{n}(Z'^\dag,Z)=(\mathbb{1}_{2N}-Z'^\dag Z)^{-\mathbb{n}},\label{Bergmandef}
\ee
where $\overline{\psi_\mathbb{n}(U')}=\det(A'^\dag)^{-\mathbb{n}}$, $Z'^\dag=(B'A'^{-1})^\dag$ and $\mathfrak{K}_\mathbb{n}(Z'^\dag,Z)$ is the so called 
\emph{Bergman kernel} of $\mathbb{D}_{4N^2}$. This suggests us to restrict ourselves to functions $\psi(U)=\phi(Z)\psi_\mathbb{n}(U)$, where 
$\phi(Z)$ denotes an arbitrary analytic holomorphic function. Since $|\psi_\mathbb{n}(U)|^2=\det(\mathbb{1}_{2N}-Z^\dag Z)^{\mathbb{n}}$, the 
scalar product \eqref{spc} on the whole group $\rmu(2N,2N)$ can be restricted to $\mathbb{D}_{4N^2}$ as
\be
\frac{(\psi|\psi')}{v(\rmu(2N))^2}c_\mathbb{n}=\int_{\mathbb{D}_{4N^2}}\overline{\phi(Z)}\phi'(Z)d\mu_\mathbb{n}^N(Z,Z^\dag)\equiv\langle \phi|\phi'\rangle,\label{scp2}
\ee
where $v(\rmu(2N))$ denotes the total volume of $\rmu(2N)$ and 
\be
d\mu_\mathbb{n}^N(Z,Z^\dag)=c_\mathbb{n}|\psi_\mathbb{n}(U)|^2\det(\mathbb{1}_{2N}-Z^\dag Z)^{-4N}|dZ|=c_\mathbb{n}\det(\mathbb{1}_{2N}-ZZ^\dag)^{\mathbb{n}-4N}|dZ|\label{measurecolorcartan}
\ee
denotes the measure on $\mathbb{D}_{4N^2}$. The constant  (the \emph{formal degree}) 
\[c_\mathbb{n}=\pi^{-4N^2}\prod_{j=1}^{2N}\frac{(\mathbb{n}-j)!}{(\mathbb{n}-2N-j)!}\] 
has been introduced to make the unit function $\phi(Z)=1$ normalized 
(see \cite{Perelomovbook,Leverrier}).

Let us consider then the Hilbert space $\mathcal{H}_\mathbb{n}(\mathbb{D}_{4N^2})$ of 
square integrable holomorphic (wave) functions $\phi(Z)$ on the phase space $\mathbb{D}_{4N^2}$ with scalar product $\langle \phi|\phi'\rangle$ 
given in terms of the measure $d\mu_\mathbb{n}^N$. Finiteness of this measure requires $\mathbb{n}\geq 4N$. 
Taking into account that $Z=BA^{-1}$, the group action $U''=U'^{-1}U$ induces a  ``colored M\"obius'' transformation of a point $Z\in \mathbb{D}_{4N^2}$ 
under a colored conformal  translation $U'$ as:
\be
Z\stackrel{U'}{\rightarrow} Z'=B''A''^{-1}=(D'^\dag Z-C'^\dag)(A'^\dag-B'^\dag Z)^{-1}.\label{Ztransf}
\ee
The regular representation $\mathcal{U}$ on $L^2(\rmu(2N,2N),d\mu)$ can then be straightforwardly projected onto 
$\mathcal{H}_\mathbb{n}(\mathbb{D}_{4N^2})$ as $[\mathcal{U}_\mathbb{n}(U)\phi](Z)\equiv [\mathcal{U}(U)\psi](U)/\psi_\mathbb{n}(U)$ for $\psi(U)=\phi(Z)\psi_\mathbb{n}(U)$. 
Let us summarize the previous construction in the following proposition.

\begin{prop} For any colored conformal transformation $U=\left(\ba{cc} A& C
\\ B &D\ea\right)\in \mathrm{U}(2N,2N)$, the following action  
\be
\phi(Z)\stackrel{U}{\rightarrow} [\mathcal{U}_\mathbb{n}(U)\phi](Z)=\det(A^\dag+B^\dag Z)^{-\mathbb{n}}
\phi\left((D^\dag Z-C^\dag)(A^\dag-B^\dag Z)^{-1}\right)\equiv \phi_U(Z),\label{uircolcomf}
\ee
defines a unitary irreducible square integrable (discrete series) representation $\mathcal{U}_\mathbb{n}$ of $\mathrm{U}(2N,2N)$ on the Hilbert space 
$\mathcal{H}_\mathbb{n}(\mathbb{D}_{4N^2})$ of analytic functions $\phi(Z)$ on the colored Cartan domain $\mathbb{D}_{4N^2}$ with integration measure 
$d\mu_\mathbb{n}^N(Z,Z^\dag)$.
\end{prop}
\noindent \textbf{Proof:} It is easy to see that $\mathcal{U}_\mathbb{n}(U)\mathcal{U}_\mathbb{n}(U')=\mathcal{U}_\mathbb{n}(UU')$ (group homomorphism). Irreducibility is related to the fact that the constant 
function $\phi(Z)=1$ is mapped to $\det(A^\dag+B^\dag Z)^{-\mathbb{n}}$ (the multiplier), which can be expanded in a complete basis of homogeneous polynomials of arbitrary homogeneity degree 
in the $4N^2$ complex entries of $Z$ (see \cite{EMSMTA} for a relation of this expansion with the MacMahon-Schwinger's Master Theorem).  
In order to prove unitarity, i.e. $\langle \phi_U|\phi_U\rangle=\langle \phi|\phi\rangle$, we can use the constraints 
$U\Gamma U^\dag=U^\dag \Gamma U=\Gamma$ to realize that the weight function 
$\det(\mathbb{1}_{2N}-ZZ^\dag)^{\mathbb{n}-4N}$ and the Lebesgue measure $|dZ|$ transform as
\[\det(\mathbb{1}_{2N}-Z'Z'^\dag)^{\mathbb{n}-4N}=|\det(A^\dag+B^\dag Z)|^{-2(\mathbb{n}-4N)}\det(\mathbb{1}_{2N}-ZZ^\dag)^{\mathbb{n}-4N}\]
and 
\[|dZ|=|dZ'| |\det(A^\dag+B^\dag Z)|^{8N},\]
respectively, for $Z'=(D^\dag Z-C^\dag)(A^\dag-B^\dag Z)^{-1}$. Therefore, the Jacobian determinant $|\det(A^\dag+B^\dag Z)|^{8N}$, 
the multipliers product $|\det(A^\dag+B^\dag Z)|^{-2\mathbb{n}}$ and the weight function factor 
$|\det(A^\dag+B^\dag Z)|^{2(\mathbb{n}-4N)}$ exactly 
compensate each other to give the isometry relation $\langle \phi_U|\phi_U\rangle=\langle \phi|\phi\rangle$ $\blacksquare$

For completeness and future use, let us provide a differential realization of the colored conformal generators \eqref{colconfalg} on holomorphic functions $\phi\in\mathcal{H}_\mathbb{n}(\mathbb{D}_{4N^2})$. 
Let us denote by $\mathcal{M}_U^\mathbb{n}(Z)=\det(A^\dag+B^\dag Z)^{-\mathbb{n}}$ the multiplier in \eqref{uircolcomf}, so that $\phi_U(Z)=\mathcal{M}_U^\mathbb{n}(Z)\phi(Z')$. 
For one-parameter ($t$) group translations $U(t)=e^{\ic t T^{\alpha a}}$ generated by a colored conformal generator $T^{\alpha a}$ in \eqref{colconfalg}, the 
associated conformal differential operator is defined by its action on holomorphic functions as
\be
\mathcal{T}^{\alpha a}\phi(Z)=-\ic\left.\frac{d\phi_U(Z)}{dt}\right|_{t=0}=-\ic\left.\frac{d\mathcal{M}_U^\mathbb{n}(Z)}{dt}\right|_{t=0}
\phi(Z) -\ic\left.\frac{d z'_{\mu b}}{dt}\right|_{t=0}\frac{\partial}{\partial z_{\mu b}}\phi(Z), \label{difgen} 
\ee
where we are using coordinates $z_{\mu b}=\frac{1}{2}\tr(Z\sigma^\mu\otimes\lambda^b)$ and the Einstein summation convention. 
For example, the differential operator associated to the colored dilation  
\[ D^{a}=D\otimes \lambda^a=\frac{1}{2}\left(\begin{array}{cc} -\sigma^0\otimes {\lambda^a} & 0 \\ 0 & \sigma^0\otimes{\lambda^a} \\ \end{array}\right)\] 
is
\be \mathcal{D}^a=-\mathbb{n}\,\tr(\lambda^a)-\frac{1}{4}\tr\left(\{\sigma^0\otimes\lambda^a,Z\}\sigma^\mu\otimes\lambda^b\right)\frac{\partial}{\partial z_{\mu b}},
\ee
where $\{A,B\}=AB+BA$ means the anticommutator. Note that, according to the nomenclature adopted in this article, $\tr(\lambda^a)=\sqrt{2N}\delta_{a,0}$ since $\lambda^a$ 
are traceless except $\lambda^0=\mathbb{1}_{2N}/\sqrt{2N}$. The differential operator associated to colored translations is
\be P^{\nu a}=P^\nu\otimes \lambda^a=\left(\begin{array}{cc} 0 & \sigma^\nu\otimes {\lambda^a} \\ 0 & 0 \\ \end{array}\right)
\to \mathcal{P}^{\nu a}=\frac{\partial}{\partial z_{\nu a}}.
\ee
For colored accelerations 
\[ K^{\nu a}=K^\nu\otimes \lambda^a=\left(\begin{array}{cc} 0 & 0 \\ \check\sigma^\nu\otimes {\lambda^a} & 0 \\ \end{array}\right)\]
we have 
\be \mathcal{K}^{\nu a}=2\mathbb{n}\, z^{\nu a}-\frac{1}{2}\tr\left(Z\check\sigma^\nu\otimes\lambda^aZ\sigma^\mu\otimes\lambda^b\right)\frac{\partial}{\partial z_{\mu b}},
\ee
where $z^{\nu a}=\frac{1}{2}\tr(Z\check{\sigma}^\nu\otimes \lambda^a)=\eta^{\nu\mu}z_{\mu a}$. The expression of colored Lorentz differential operators  
$\mathcal{M}^{\mu\nu a}$ is a bit more bulky. Their aspect gets simpler for $N=1$, acquiring the more familiar form
\be
\mathcal{D}=-\mathbb{n}-z_\mu\partial^\mu,\; \mathcal{P}^\mu=\partial^\mu,\, \mathcal{K}^\mu=-z^2\mathcal{P}^\mu-2z^\mu\mathcal{D}, \; \mathcal{M}^{\mu\nu}=z^\mu\partial^\nu-z^\nu\partial^\mu,
\ee
where $\partial^\mu=\partial/\partial z_\mu$ and $z^2=\eta^{\mu\nu}z_\mu z_\nu=z^\nu z_\nu$. Note that commutation relations get unaltered under the conformal ``Born reciprocity principle'' 
\cite{spinning} $D\leftrightarrow -D$, $P_\mu\leftrightarrow K_\mu$. 

\section{Twistor nonlinear sigma models for conformal massive colored particles}\label{twistnlsm}

In this section we shall formulate nonlinear sigma models on the Grassmannian $\mathbb{D}_{4N^2}$, either as a configuration space or a phase space of a 
colored conformal massive particle, in terms of colored twistors (``ctwistors'' for short). 
Let us define $M$-ctwistors  $\zeta$ as a juxtaposition of $M$ (column) linearly independent 1-ctwistors  
$\zeta_j=(a_{1,j},\dots,a_{2N,j},b_{1,j},\dots,b_{2N,j})^t\in \mathbb{C}^{4N}, j=1,\dots,M$, that is: 
\be
\mathbf{\zeta}=\bordermatrix{&\zeta_1 & \dots &\zeta_M  \cr & a_{11}  &\dots & a_{1,M}  \cr & \vdots  & 
& \vdots \cr & a_{2N,1} &  \dots & a_{2N,M}
\cr & b_{11}  &\dots & b_{1,M}  \cr & \vdots  & 
& \vdots \cr & b_{2N,1} &  \dots & b_{2N,M}}.
\label{Zbf}
\ee
Here ctwistors $\zeta_j$ are coordinatized by a pair of $2N$-component colored spinors $a$ and $b$, resembling standard two-component ``undotted and dotted'' 
(or chiral and anti-chiral) spinors, in van der Waerden notation, respectively {(see later on Remark \ref{incidence} for more information 
about the traditional van der Waerden notation and the generalization of Penrose incidence relations)}. 
For $N=1$, we know \cite{Todorov} that  $\mathbb{C}^{4}$ 1-twistors  describe massless particles, whereas $2$-twistor compounds 
describe conformal massive particles. For standard 1-twistors, the quantity $\zeta^\dag \Gamma \zeta$ is $\mathrm{U}(2,2)$ invariant and represents the helicity $s$ of the corresponding 
massless particle. We are interested in the massive case, therefore, we shall set $M=2N$ in the sequel. $2N$-ctwistors are 
subject to the constraint $\zeta^\dag \Gamma \zeta=\kappa\mathbb{1}_{2N}$, where  $\mathrm{sgn}(\kappa)=\pm 1$  makes reference to the two open orbits 
$\mathbb{D}_{4N^2}^+$ and $\mathbb{D}_{4N^2}^-$ that the complex pseudo-Grassmannian manifold 
$\mathbb{D}_{4N^2}$ of $2N$-planes in $\mathbb{C}^{4N^2}$ carries. The quantum counterpart of this constraint will fix the value of the conformal scale $\mathbb{n}$ [see later on Eq. \eqref{qconst}]. 
This constraint is preserved by transformations $\zeta\to U\zeta V$, with $U\in\mathrm{U}(2N,2N)$ and $V\in \mathrm{U}(2N)$. More explicitly
\be
\zeta=\left(\ba{c} A \\ B\ea\right)\to U\zeta V=\left(\ba{cc} A'& C'
\\ B' &D'\ea\right)\left(\ba{c} A \\ B\ea\right) V=\left(\ba{c} (A'A+C'B)V \\ (B'A+D'B)V\ea\right).
\ee

Let us motivate the appearance of $\mathbb{D}_{4N^2}$ from constrained Lagrangian mechanics of $2N$-ctwistors $\zeta$, 
either as a configuration or a phase space of a massive colored conformal particle. 

\subsection{U($2N$) gauge twistor model and geodesic motion on $\mathbb{D}_{4N^2}$}

We make $\zeta$ depend on time $\zeta(t)$ and start considering the Lagrangian
\be
\tilde{L}=\tr(\frac{d\zeta^\dag}{dt}\Gamma\frac{d\zeta}{dt}).\label{twistL}
\ee
This Lagrangian is invariant under rigid transformations $\zeta\to U\zeta V$, $U\in \mathrm{U}(2N,2N), V\in \mathrm{U}(2N)$. Now we want to promote rigid right transformations 
$V\in \mathrm{U}(2N)$ to local (gauge) transformations $V(t)$ depending arbitrarily on time $t$. This means that the configuration space is restricted to the Grassmannian space 
of all $2N$-dimensional linear subspaces of $\mathbb C^{4N}$. The connection with the coset $\mathbb{D}_{4N^2}=\rmu(2N,2N)/\rmu(2N)^2$ 
is the following. We can think of colored conformal transformations  
\be U=\left(\ba{cc} A& C\\ B &D\ea\right)=(\zeta|\zeta_\perp), \quad \zeta=\left(\ba{c} A \\ B\ea\right), \; \zeta_\perp=\left(\ba{c} C \\ D\ea\right),\ee
as a juxtaposition of two perpendicular ($\zeta^\dag\Gamma\zeta_\perp=0$) $2N$-ctwistors  $\zeta$ and $\zeta_\perp$ fulfilling the constraint 
$\zeta^\dag \Gamma \zeta=\mathbb{1}_{2N}$ and $\zeta^\dag_\perp \Gamma \zeta_\perp=-\mathbb{1}_{2N}$, for  $\Gamma$ with signature $(1,\stackrel{2N}{\dots},1,-1,\stackrel{2N}{\dots},-1)$. 
We make this choice for the sake of convenience.

The coset representative \eqref{cosetcoord} obtained from the Iwasawa decomposition \eqref{Iwasawa} then provides a gauge fixing given by
\be  \zeta=\left(\ba{c}  \mathbb{1}_{2N}\\  Z \ea\right) (\mathbb{1}_{2N}- Z^\dag Z)^{-1/2}, \quad Z=BA^{-1}\label{gaugefix}
\ee
Therefore, the $2N$-ctwistor $\zeta$ is no longer a set of $2N$ independent ctwistors $\zeta_j$, but only carries $8N^2$ real degrees of freedom 
[the $4N^2$ complex entries $z_{ij}$ of $Z$].

\begin{rem}\label{incidence}{ \textbf{(Gauge Fixing and Colored Penrose Incidence Equations)} As promised, let us make an brief aside 
about the relation between our notation and the traditional van der Waerden notation, together with an interpretation  of the gauge fixing 
\eqref{gaugefix}  as a generalization of the traditional Penrose incidence relations. A two-twistor compound describing a conformal massive particle $(N=1)$ 
is a juxtaposition $\zeta=(\zeta_1\vert\zeta_2)$ of two (column) 1-twistors  $\zeta_1=(\bar\pi_{\dot{\alpha}},\omega^\alpha)^t$ and 
$\zeta_2=(\bar\eta_{\dot{\alpha}},\lambda^\alpha)^t$, which in turn  are defined by a pair of complex Weyl spinors  (dotted and undotted spinor 
indices correspond to positive and negative chirality; we are using the notation of, for example, Ref. \cite{PRD73}). Arranging the two-twistor compound as:}
\begin{equation}
{\zeta=(\zeta_1\vert\zeta_2)=\left(\ba{c} A \\ B\ea\right),\; 
A=\left(\ba{cc}\bar\pi_{\dot{1}} & \bar\eta_{\dot{1}} \\  \bar\pi_{\dot{2}} & \bar\eta_{\dot{2}}\ea\right), 
B=\left(\ba{cc}\omega^1 & \lambda^1\\ \omega^2 & \lambda^2\ea\right),  }
\end{equation}
{the gauge fixing $Z=BA^{-1}$ in \eqref{gaugefix} straightforwardly provides the composite complex Minkowski coordinates $z_\mu=\um\tr(Z\sigma^\mu)$ 
described by the well-known Penrose formula}
\begin{equation}
{Z=BA^{-1}\to z^{\alpha\dot{\beta}}=\frac{\omega^\alpha\bar\eta^{\dot{\beta}}-\lambda^\alpha\bar\pi^{\dot{\beta}}}{\det(A)},\; 
\det(A)=\bar\pi^{\dot{\alpha}}\bar\eta_{\dot{\alpha}}\not=0,  }
\end{equation}
{where we rise and lower dotted and undotted indices with the two-dimensional Levi-Civita symbol, 
namely $\bar\pi^{\dot{\alpha}}=\epsilon^{\dot{\alpha}\dot{\beta}}\bar\pi_{\dot{\beta}}$. The Penrose incidence relations simply state that}
\begin{equation}
{B=ZA\Leftrightarrow \omega^\alpha=z^{\alpha\dot{\beta}}\bar\pi_{\dot{\beta}},\, \lambda^\alpha=z^{\alpha\dot{\beta}}\bar\eta_{\dot{\beta}}. }
\end{equation}
{Therefore, using our compact matrix notation, Penrose formula and incidence relations are strightforwardly  generalized to the colored 
($N>1$) case simply as $Z=BA^{-1}$ and $B=ZA$, respectively. The Grassmannian space  $\mathbb{D}_{4N^2}=\rmu(2N,2N)/\rmu(2N)^2$ 
of all $2N$-dimensional linear subspaces of $\mathbb C^{4N}$ (the so called null-planes for $N=1$) are parametrized by the composite colored complex Minkowski coordinates $z_{\mu a}$ in \eqref{ccc}. 
Once stated the relation to standard spinorial geometry, we shall rather continue with our compact matrix notation. $\blacksquare$} 
\end{rem}

Although we want $\zeta(t)$ and $\zeta(t) V(t)$ to be (gauge) equivalent, the Lagrangian \eqref{twistL} is not $\mathrm{U}(2N)$-gauge invariant. 
In order to make it gauge invariant, we must perform minimal coupling, replacing  ${d\zeta}/{dt}$ by the covariant derivative 
${D\zeta}/{dt}={d\zeta}/{dt}-\ic\zeta \mathcal A$, with $\mathcal A$ a $\mathrm{U}(2N)$-gauge potential transforming as $\mathcal A\to V^\dag \mathcal{A} V-\ic V^\dag dV/dt$. 
Therefore, the Lagrangian 
\be
L=\tr[ \left(\frac{D\zeta}{dt}\right)^\dag \Gamma\frac{D\zeta}{dt}],\label{twistLgi}
\ee
becomes $\mathrm{U}(2N)$-gauge invariant. From the equations of motion, and given that $\mathcal A$ is an auxiliary (not a dynamical) field,  
we obtain $\mathcal A=-\ic\zeta^\dag \Gamma {d\zeta}/{dt}=\ic{d\zeta^\dag}/{dt}\Gamma\zeta$. Using this, the covariant derivative 
can be written as $D\zeta/dt=P(\zeta,\zeta^\dag)d\zeta/dt$ with $P(\zeta,\zeta^\dag)=(\mathbb{1}_{2N}-\zeta\zeta^\dag\Gamma)$ a $2N\times 2N$ 
projector fulfilling  $P\zeta=0$ (it projects onto the orthogonal subspace to $\zeta$ spanned by  $\zeta_\perp$) and $P^\dag=\Gamma P\Gamma, P^\dag\Gamma P=\Gamma P$. With this 
information, the Lagrangian  \eqref{twistLgi} 
can be equivalently written as
\be
L=\tr\left[ \frac{d\zeta^\dag}{dt} \Gamma P(\zeta,\zeta^\dag)\frac{d\zeta}{dt}\right]=
\frac{1}{2}\tr\left(\frac{d\mathfrak{T}}{dt}\frac{d\mathfrak{T}}{dt}\right), \quad \mathfrak{T}=\zeta\zeta^\dag\Gamma,\label{twistLgi2}
\ee
where we have used that $\frac{d}{dt}(\zeta^\dag\Gamma\zeta)=0$ to derive the right-hand part of \eqref{twistLgi2}. The last expression shows the Lagrangian 
written in terms of the gauge invariant quantities $\mathfrak{T}$. Using a basis $\{\tilde\lambda^a=\Gamma\lambda^a, a=0,\dots,16N^2-1\}$ of 
$2N\times 2N$ pseudo-hermitian matrices $\tilde\lambda^a$ of the colored conformal Lie algebra $\rmuu(2N,2N)$, 
with $\lambda^a$ the generalized $\rmu(4N)$ Pauli matrices defined after \eqref{PauliN}, 
we can expand $\mathfrak{T}$ in this basis with coefficients (the generalized Hopf map)
\be
\mathfrak{T}=\mathfrak{T}_a\tilde\lambda^a, \quad \mathfrak{T}^a=\tr(\mathfrak{T}\tilde\lambda^a)=\tr(\zeta\zeta^\dag\lambda^a), \; a=1,\dots,16N^2-1,\label{sunspin}
\ee
where $\mathfrak{T}^a=\tilde\delta^{ab}\mathfrak{T}_b$ and $\tilde\delta^{ab}$ denotes the  $\rmu(2N,2N)$ Killing form 
\be
\tilde\delta^{ab}=\tr(\tilde\lambda^a\tilde\lambda^b)=\tr(\lambda^a\Gamma\lambda^b\Gamma)=\mathrm{diag}(1,\stackrel{8N^2}{\dots},1,-1,\stackrel{8N^2}{\dots},-1)
\ee
in a specific ordering [remember that we have $\tr(\lambda^a\lambda^b)=\delta^{ab}$].  Note that the gauge invariant quantities $\mathfrak{T}_a$ are constrained by
\be\tr(\mathfrak{T}\mathfrak{T})= \tr(\zeta\zeta^\dag\Gamma\zeta\zeta^\dag\Gamma)=\tr(\mathbb{1}_{2N}) = 2N =\mathfrak{T}_a\tilde\delta^{ab}\mathfrak{T}_b ,\label{herm}
\ee
where we have used the constraint $\zeta^\dag\Gamma\zeta=\mathbb{1}_{2N}$. Therefore, the Lagrangian \eqref{twistLgi2} can also be written as 
\be
L=\frac{1}{2} \frac{d\mathfrak{T}_a}{dt}\tilde\delta^{ab}\frac{d\mathfrak{T}_b}{dt}.\label{sunspinL}
\ee
This expression resembles the usual Lagrangian for a free particle on the $M$-sphere $\mathbb{S}^M$,  $L=\frac{1}{2} \dot{x}_a\delta^{ab}\dot{x}_a$, with the constraint 
${x}_a\delta^{ab}x_b=R^2$ (the squared radius), replacing the Euclidean $\delta$ by the non-Euclidean $\tilde\delta$ metric. Therefore,  the Lagrangian \eqref{twistLgi2} 
describes geodesic motion on $\mathbb{D}_{4N^2}$. 
Another equivalent expression for  \eqref{twistLgi2} is given in terms of the minimal coordinates $Z\in \mathbb{D}_{4N^2}$ though the gauge fixing \eqref{gaugefix} as
\be
L= \tr \left(  \Delta_2^2 \frac{d Z}{dt} \Delta_1^2  \frac{d Z^\dag}{dt}  \right),
\ee
where we have used that $\Delta_2^2 Z =  Z \Delta_1^2$ with $\Delta_{1,2}$ in \eqref{cosetcoord}.  

{Note that, the colored conformal domain  $\mathbb{D}_{4N^2}$ here plays the role of a configuration space, rather than a phase space. For $N=1$, the eight-dimensional 
conformal domain $\mathbb{D}_{4}$ has been propossed in \cite{Jadczyk} to replace the four-dimensional Minkowski space-time at short distances (at the microscale or high energies). 
This approach has to do with Born's reciprocity principle \cite{Born1,Born2}, which conjectures that the basic underlying physical space is the eight-dimensional $\{x_\mu,p_\mu\}$ phase space and 
replaces the Poincar\'e line element $ds^2=dx_\mu d x^\mu$ by the Finslerian $d\tilde{s}^2=dx_\mu d x^\mu+\ell^4_{\mathrm{min}}dp_\mu d p^\mu/\hbar^2$, where $\ell_{\mathrm{min}}$ is a minimal length 
(maximal momentum transfer). Interesting physical phenomena like dark matter and black hole cosmology have been reinterpreted as an effect of Born's reciprocal relativity theory \cite{castro}. 
The adaptation of Born's reciprocity principle to conformal relativity was also put forward in \cite{spinning}, and an explanation of the Unruh effect (vacuum radiation in accelerated frames) 
as a spontaneous breakdown of the conformal symmetry $\rmu(2,2)$  was given in \cite{Unruh1,Unruh2}. }

\subsection{Berry Lagrangian and colored conformal massive particles}

In this article we are more interested in the colored complex Minkowski space $\mathbb{D}_{4N^2}$ as a phase space, rather than a configuration space. This construction 
can also be achieved by considering $\mathrm{U}(2N)$-gauge invariant ctwistor Lagrangians, but this time linear in $d\zeta/dt$  (that is, singular Lagrangians) 
in a Berry-like form
\be
{L}_B=  \ic\, \tr(\zeta^\dag\Gamma\frac{d\zeta}{dt})=-\tr(\mathcal A).\label{twistLgi3}\ee
One can prove that \eqref{twistLgi3} is also invariant under rigid colored conformal transformations,  
$\zeta\to U\zeta, U\in \mathrm{U}(2N,2N)$,  and semi-invariant under local (gauge) transformations $\zeta(t)\to \zeta(t) V(t), V(t)\in \mathrm{U}(2N)$, up to a total derivative 
\be
{L}_B\to {L}_B+\ic\, \frac{d\,\tr(\ln V)}{dt}.
\ee
This Lagrangian encodes the basic Poisson brackets.  The fact that $\partial {L}_B/\partial\dot\zeta=\ic\zeta^\dag\Gamma$ implies that  
conjugate momenta of $\zeta$ are $\pi=\ic\zeta^\dag\Gamma$, whereas $\pi^\dag=\partial {L}_B/\partial\dot\zeta^\dag=0$. Singular Lagrangians can be approached by 
Dirac formalism, treating $\pi^\dag=0$ as a constraint. We shall not enter into details and we will just state the essential Poisson brackets among phase space variables 
$\zeta=(\zeta_{ij})$ and $\zeta^\dag=(\bar\zeta_{ji})$  as
\be \{\zeta_{ij},\pi_{kl}\}=\delta_{ik}\delta_{jl}, \rightarrow \{\zeta_{ij},\bar\zeta_{kl}\}=-\ic\delta_{ik}\Gamma_{jl},\ee
which endow  ctwistors  with a canonical symplectic structure. Under these Poisson brackets, the conserved quantities $\mathfrak{T}^a$ in \eqref{sunspin}, associated 
to the colored conformal symmetry, close the $\mathrm{u}(2N,2N)$ Lie algebra commutation relations 
\be
\{\mathfrak{T}^a,\mathfrak{T}^b\}=\tr(\mathfrak{T}[\tilde{\lambda}^a,\tilde{\lambda}^b]).
\ee

The standard quantization mapping $\zeta\to\hat{\zeta}$ assigns bosonic 
annihilation and creation quantum operators to the classical $\zeta$ and $\bar \zeta$ phase space quantities. However, due to the indefinite character of $\Gamma$, we have to split 
$\hat\zeta$ in the following form. Actually, a direct application of the standard canonical quantization rules states that
\be
[\hat\zeta_{ij},\hat\zeta^\dag_{j'i'}]=\ic\widehat{\{\zeta_{ij},\bar \zeta_{i'j'}\}}=\delta_{ii'}\Gamma_{jj'}.
\ee
{Traditionally, gauge invariance and indefinite metrics create dificulties in the quantization process (ghosts, negative energy, etc). In our case, the solution 
resembles the original  Dirac's idea of reinterpreting  some annihilation operators of particles as creation operators of holes. In our case we see that commutations relations for degrees of freedom 
$\hat\zeta_{ij}$ with $j> 2N$ have de reverse sign as desired for standard bosonic commutation relations are $[\hat a,\hat a^\dag]=1$. The solution is to let 
 $\hat\zeta_{ij}$ represent annihilation operators $\hat a$ for $j\leq 2N$, whereas 
$\hat\zeta_{ij}$ represent creation operators $\hat b^\dag$ for $j> 2N$.} Otherwise stated, to set the quantization mapping as 
\be
\hat\zeta=\begin{pmatrix}
            \hat{A}\\ \hat{B}^\dag
           \end{pmatrix},\quad \hat{A}=(\hat a_{ij}),\; \hat{B}=(\hat b_{ij}),\; i,j=1,\dots,2N.\label{calzeta}
\ee
This is just a convention, as we  could also have arrived to the choice $\hat\zeta^\dag =(\hat{A}\; \hat{B}^\dag)$ by changing the sign of the Lagrangian. 
Compare this splitting of $\hat\zeta$ with the separation of the classical ctwistor \eqref{Zbf} into a pair of colored conjugate (dotted and undotted) spinors $a$ and $b$. 
{What we are saying is that undotted spinors have a ``hole'' nature (with this convention) in the quantization process. Extra constraints have to be imposed [see later on Eq. \eqref{qconst}] 
stating that, in particular, the total excess of particle over hole quanta must be fixed.}
Commutation relations are invariant under colored conformal group transformations $\hat\zeta\to U\hat\zeta, \, U\in \rmu(2N,2N)$. {In fact, this quantization mapping is 
closely related to the Jordan-Schwinger oscillator representation of the pseudo-unitary u$(p,q)$ Lie algebra generators [see later on Eq. \eqref{bosrepresym}]}. 

\section{Bosonic representation of a colored conformal massive quantum particle}\label{bosonsec}

Let us construct the Hilbert space and the basic observables of a colored conformal massive quantum particle in a different (Fock space) picture to the holomorphic 
representation offered at the end of section \ref{colorconf}. Both pictures are related by the coherent state representation of a given quantum state (see later in this section). 

We have two kinds of quanta: $a$-type and $b$-type; due to their conjugated nature inside $\hat\zeta$, we shall refer to them sometimes as ``particle and hole quanta'', respectively. 
The Fock space is generated from a normalized vacuum $|0\rangle$ (annihilated by $\hat a_{ij}$ and $\hat b_{ij}$) by repeated action 
of creation operators
\be
|n^a\ra \otimes |{n}^b\ra=\prod_{i,j=1}^{2N}
\frac{(\hat a^\dag_{ij})^{n_{ij}^a}(\hat b^\dag_{ij})^{n_{ij}^b}}{\sqrt{n_{ij}^a!n_{ij}^b!}}|0\ra,\label{grassmannbasis2}
\ee
where $n^{a,b}$ denote $2N\times 2N$ matrices with integer entries $n_{ij}^{a,b}$ (the occupancy numbers of $a$- and $b$-type bosons). 
A unitary representation of the colored conformal group in Fock space is then given by the 
oscillator (Jordan-Schwinger) realization of the $16N^2$ colored conformal generators $T^{\alpha a}$ in \eqref{colconfalg}, given  by
\be
\hat{T}^{\alpha a}=-\tr(\hat\zeta^\dag \Gamma T^{\alpha a} \hat\zeta),\label{bosrepresym} \ee
which is the quantum counterpart of the classical Hopf map \eqref{sunspin}. This oscillator realization is in fact extensible to general $U(p,q)$ 
\cite{Todorov2,couplingUpq} and it became popular after \cite{u66},  
who discussed the use of $\rmu(6,6)$ to classify hadrons; in this case barions and antibarions belong to mutually conjugate representations 
with respect to $\rmu(6)$. Using commutation relations for creation and annihilation operators, one can verify that 
$[\hat{T}^{\alpha a},\hat{T}^{\beta b}]=-\tr(\hat\zeta^\dag\Gamma[{T}^{\alpha a},{T}^{\beta b}]\hat\zeta)$, which implies that \eqref{bosrepresym} 
defines a Lie algebra representation of $u(2N,2N)$ in the Fock space defined by \eqref{grassmannbasis2}. All colored conformal generators $\hat{T}^{\alpha a}$ commute with 
$\hat{T}^{00}=\tr(\hat\zeta^\dag \Gamma \hat\zeta)$ (the linear Casimir) and, therefore, the representation is reducible. In order to reduce it, we have to fix the value of 
\be
\hat\zeta^\dag \Gamma \hat\zeta=\hat{A}^\dag\hat{A}-\hat{B}\hat{B}^\dag=2Nc \mathbb{1}_{2N},\label{qconst}
\ee
which is the quantum counterpart of the ctwistor constraint $\zeta^\dag\Gamma\zeta=\kappa\mathbb{1}_{2N}$  (the helicity for 1-twistors) defined after \eqref{Zbf}. The value 
of the constant $c$ will be related to the conformal or scale dimension $\mathbb{n}$, introduced after \eqref{spc}, and the number of colors $N$ in Proposition 
\ref{groundstate}. Taking into account that $\hat{B}\hat{B}^\dag=\hat{B}^\dag\hat{B}+2N\mathbb{1}_{2N}$ and the constraint \eqref{qconst}, the linear Casimir is
\be 
\hat{T}^{00}=-\tr(\hat\zeta^\dag \Gamma \hat\zeta)=4N^2-\sum_{i,j=1}^{2N} \hat{n}^a_{ij}-\hat{n}^b_{ij}=-4N^2c,
\ee
which says that the total excess of particle over hole quanta is always a fixed quantity $n^a-n^b=4N^2(c+1)$. The constant $c$ is chosen so that the excess 
$n^a-n^b$ is an integer. Therefore, particle and hole quanta must be created and annihilated 
by pairs (``excitons''). Actually, the oscillator representation of the colored dilation operator $D$ in \eqref{confalgamma}
 \be
 D^0=\frac{1}{2}\left(\ba{cc} -\sigma^0 & 0\\ 0 &\sigma^0 \ea\right)\otimes\lambda^0\rightarrow \hat{D}^0=\frac{1}{2\sqrt{2N}}\tr(\zeta^\dag\zeta)=
 \frac{1}{2\sqrt{2N}}\tr(\hat{A}^\dag\hat{A}+\hat{B}\hat{B}^\dag),
 \ee
measures the total number $(n^a+n^b)/2$ of excitons (particle-hole pair quanta $ab$). The oscillator representation of colored accelerations
\be
K^{\nu a}=\left(\begin{array}{cc} 0 & 0 \\ \check\sigma^\nu\otimes {\lambda^a} & 0 \\ \end{array}\right)\rightarrow \hat K^{\nu a}=\tr(\hat B \check\sigma^\nu\otimes {\lambda^a}\hat A  )
\ee
annihilates excitons, whereas the oscillator representation of colored translations 
\be
P^{\nu a}=\left(\begin{array}{cc} 0 & \sigma^\nu\otimes {\lambda^a} \\ 0 & 0 \\ \end{array}\right)\rightarrow \hat P^{\nu a}=-\tr(\hat A^\dag\sigma^\nu\otimes {\lambda^a}\hat B^\dag)
\ee
creates excitons.  Excitons are not exactly bosons, since the basic commutations relations between creation and annihilation operators of excitons, $\lb {\hat{K}}^\mu,{\hat{P}}^\nu\rb = 
2(\eta^{\mu\nu}{\hat{D}}+{\hat{M}}^{\mu\nu})$, include corrections in the number of particle-hole pairs arising from the interaction between excitons.

We can choose the ground state $|\psi_\mathbb{n}\rangle$ to be made of either particle $(a)$ or hole $(b)$ quanta, leading to two different (inequivalent) 
representations. Let us choose particle quanta this time for the content of our ground state (the other option is similar). The structure of the ground state $|\psi_\mathbb{n}\rangle$ 
is the Fock space counterpart of the lowest-weight state $\psi_\mathbb{n}(U)=\det(A)^{-\mathbb{n}}$ defined after \eqref{spc}. Its structure is described in the 
following Proposition.
\begin{prop}\label{groundstate}  Let $\mathbb{n}=2N(c+1)$, with $c>-1$, an integer. The Fock state
\be
|\psi_\mathbb{n}\ra=\frac{\det(\hat A^\dag)^{\mathbb{n}}}{\mathcal{N}_{\mathbb{n},N}^{1/2}}|0\ra,\quad \mathcal{N}_{\mathbb{n},N}=\prod_{q=1}^{2N} (q)_\mathbb{n}, \label{lowestweight}\ee
is made of $2N\mathbb{n}=4N^2(c+1)$ particle ($a$-type) quanta and  fulfills the
constraints $\hat\zeta^\dag \Gamma \hat\zeta|\psi_\mathbb{n}\rangle =2Nc\mathbb{1}_{2N}|\psi_\mathbb{n}\rangle$ given in \eqref{qconst}. $\mathcal{N}_{\mathbb{n},N}$ is a normalization constant, 
such that $\langle \psi_\mathbb{n}|\psi_\mathbb{n}\rangle=1$, given in terms 
of the Pochhammer symbol $(q)_\mathbb{n}=q(q+1)\dots (q+\mathbb{n}-1)$. 
Therefore, $|\psi_\mathbb{n}\rangle$ can be taken as the ground (lowest weight) state of a representation of the colored conformal group.
\end{prop}
\noindent \textbf{Proof:} On the one hand, looking at the structure of 
\[\det(\hat A^\dag)=\sum_{\sigma\in S_{2N}}\mathrm{sgn}(\sigma)\prod_{j=1}^{2N} \hat a_{j,\sigma_j}^\dag=\sum_{i_1,\dots,i_{2N}=1}^{2N}\varepsilon_{i_1,\dots,i_{2N}}\prod_{j=1}^{2N} \hat a_{j,i_j}^\dag,\]
where $S_{2N}$ is the symmetric group of degree $2N$ and $\varepsilon$ is the Levi-Civita symbol, it is clear that $\det(\hat A^\dag)^{\mathbb{n}}|0\rangle$ is made of $4N\times \mathbb{n}=4N^2(c-1)$ particle quanta. 
On the other hand, 
the basic boson commutation relations $[\hat a,\hat a^\dag]=1$ imply that $[\hat a,F(\hat a^\dag)]=F'(\hat a^\dag)$ or $\hat aF(\hat a^\dag)|0\ra=F'(\hat a^\dag)|0\ra$, 
where $F'$ denotes the formal derivative 
with respect to the argument. Let us simply write $\hat a^\dag=a^\dag$ and $\hat a=\partial/\partial a^\dag$. Therefore
\bea(\hat{A}^\dag\hat{A})_{ij}\det(\hat A^\dag)^{\mathbb{n}}|0\ra&=&
\sum_{k=1}^{2N}a^\dag_{ki}\frac{\partial}{\partial a^\dag_{kj}}\det(\hat A^\dag)^{\mathbb{n}}|0\ra\nonumber \\ 
&=& \mathbb{n}\det(\hat A^\dag)^{\mathbb{n}-1}\sum_{k=1}^{2N}a^\dag_{ki}\frac{\partial}{\partial a^\dag_{kj}}\det(\hat A^\dag)|0\ra.\nonumber
\eea
Antisymmetry implies that
\[\sum_{k=1}^{2N}a^\dag_{ki}\frac{\partial}{\partial a^\dag_{kj}}\det(\hat A^\dag)=\delta_{ij}\det(\hat A^\dag).\]
Therefore, $\hat{A}^\dag\hat{A}|\psi_\mathbb{n}\ra=\mathbb{n}\mathbb{1}_{2N}|\psi_\mathbb{n}\ra$ and $|\psi_\mathbb{n}\ra$  fulfills the  constraint \eqref{qconst}, that is
\[ \hat\zeta^\dag \Gamma \hat\zeta|\psi_\mathbb{n}\ra  =(\hat{A}^\dag\hat{A}-\hat{B}^\dag\hat{B}-2N\mathbb{1}_{2N})|\psi_\mathbb{n}\rangle=2Nc\mathbb{1}_{2N}|\psi_\mathbb{n}\rangle. \]
It remains to prove that the norm of $\det(\hat A^\dag)^{\mathbb{n}}|0\rangle$ is given by the quantity $\mathcal{N}_{\mathbb{n},N}^{1/2}$ in \eqref{lowestweight}. We proceed by mathematical induction. 
Firstly we prove that $\mathcal{N}_{1,N}=(2N)!$. Indeed, 
\[\langle 0|\det(\hat A)\det(\hat A^\dag)|0\rangle=\sum_{\sigma\in S_{2N}} 1=(2N)!.\]
Now we assume that $\langle 0|\det(\hat A)^\mathbb{n}\det(\hat A^\dag)^\mathbb{n}|0\rangle=\mathcal{N}_{\mathbb{n},N}$ and we shall prove that 
\[\langle 0|\det(\hat A)^{\mathbb{n}+1}\det(\hat A^\dag)^{\mathbb{n}+1}|0\rangle=\mathcal{N}_{\mathbb{n}+1,N}.\]
Indeed, it can be shown that 
\[\langle 0|\det(\hat A)^{\mathbb{n}+1}\det(\hat A^\dag)^{\mathbb{n}+1}|0\rangle=(\mathbb{n}+1)_{2N}\langle 0|\det(\hat A)^{\mathbb{n}}\det(\hat A^\dag)^{\mathbb{n}}|0\rangle.\]
The proof is a bit clumsy and we shall restrict ourselves to the more maneuverable $N=1$ case, which grasps the essence of the general case. In fact,
\bea \det(\hat A)\det(\hat A^\dag)^{\mathbb{n}+1}|0\rangle&=&
\left(\frac{\partial}{\partial a^\dag_{11}}\frac{\partial}{\partial a^\dag_{22}}-\frac{\partial}{\partial a^\dag_{12}}\frac{\partial}{\partial a^\dag_{21}}\right)
(a^\dag_{11}a^\dag_{22}-a^\dag_{12}a^\dag_{21})^{\mathbb{n}+1}|0\rangle\nonumber\\ &=&(\mathbb{n}+1)\mathbb{n}\det(\hat A^\dag)^\mathbb{n}|0\rangle+2(\mathbb{n}+1)\det(\hat A^\dag)^\mathbb{n}|0\rangle\nonumber\\ 
&=&(\mathbb{n}+1)_2\det(\hat A^\dag)^{\mathbb{n}}|0\rangle.
\eea
In general 
\[\det(\hat A)\det(\hat A^\dag)^{\mathbb{n}+1}|0\rangle=(\mathbb{n}+1)_{2N}\det(\hat A^\dag)^{\mathbb{n}}|0\rangle.\] 
It remains to realize that $(\mathbb{n}+1)_{2N}\mathcal{N}_{\mathbb{n},N}=\mathcal{N}_{\mathbb{n}+1,N}$ for 
$\mathcal{N}_{\mathbb{n},N}=\prod_{n=1}^{2N} (n)_\mathbb{n}$, which concludes the proof by induction $\blacksquare$

\begin{rem}
Note that the determinant structure of $|\psi_\mathbb{n}\rangle\propto \det(\hat A^\dag)^{\mathbb{n}}|0\ra$ denotes a fermion compound structure for our colored conformal 
massive particle, as made of $2N$ more elementary (massless) colored particles obeying the Pauli exclusion principle. 
However, the statistics of the compound depends on the parity of $\mathbb{n}$. {In fact, under an interchange of two columns 
(two 1-ctwistors representing two massless colored constituents) of the $2N$-ctwistor  massive 
compound $\hat{\zeta}$ in \eqref{calzeta}, the determinant $\det(\hat A^\dag)^{\mathbb{n}}$ acquires a phase $(-1)^{\mathbb{n}}$; therefore, for odd ${\mathbb{n}}$ the compound has a fermionic nature, 
whereas for even ${\mathbb{n}}$,  the compound is bosonic. That is, the conformal dimension ${\mathbb{n}}$ determines the statistics of the compound. 
This fact resembles the statistical transmutation that electrons suffer in some condensed matter systems, like fractional quantum Hall effect, when magnetic flux quanta are attached to them, thus 
forming ``composite fermions'' \cite{Jainbook}$\blacksquare$}
\end{rem}

A step by step repeated  application of exciton, creation $\hat P=- \hat B^\dag\hat A^\dag$  and annihilation $\hat K=\hat A\hat B$, ladder operators on the 
lowest-weight state $|\psi_\mathbb{n}\rangle$ provides the remainder (infinite) 
quantum states $|\psi\rangle$ of our Hilbert space $\mathcal{H}_\mathbb{n}$. This construction is rather cumbersome and has been achieved in \cite{JMP55} for $N=1$. We shall not further pursuit it here but, instead, 
we shall provide the connection between this oscillator representation and the holomorphic picture showed in section \ref{colorconf} through the introduction of 
colored conformal coherent states. We shall also show that the integer $\mathbb{n}$ defined in Proposition \ref{groundstate} coincides with the colored scale or conformal dimension 
defined in section \ref{colorconf}. 

For the Schwinger boson realization \eqref{bosrepresym} of colored conformal operators $\hat{T}^{\alpha a}$, the exponential 
$\hat{\mathcal{U}}(U)=\exp(w_{\alpha a}\hat{T}^{\alpha a})$ defines a unitary representation of $\rmu(2N,2N)$ in Fock space, with 
$U=\exp(w_{\alpha a}{T}^{\alpha a})\in \rmu(2N,2N)$ a colored conformal transformation with  matrix generators ${T}^{\alpha a}$ in \eqref{colconfalg}. 
The adjoint action of  $\hat{\mathcal{U}}(U)$ on $\hat \zeta$ and $\hat\zeta^\dag$ is simply
\be
\hat{\mathcal{U}}(U)\hat\zeta\hat{\mathcal{U}}^\dag(U)=U\hat\zeta,\quad \hat{\mathcal{U}}(U)\hat\zeta^\dag\hat{\mathcal{U}}^\dag(U)=\hat\zeta^\dag U^\dag.\label{adjointAB}
\ee
Let us introduce coherent states for the colored conformal massive particle.

\begin{defn}\label{defcoh}  For any $U\in \rmu(2N,2N)$, we define generalized coherent states 
\be
|\psi_\mathbb{n}^U\rangle=\hat{\mathcal{U}}(U)|\psi_\mathbb{n}\rangle
\ee
as Fock states $|\psi_\mathbb{n}^U\rangle$ in the orbit of the ground state $|\psi_\mathbb{n}\rangle$ under the action $\hat{\mathcal{U}}(U)$. 
\end{defn}
From \eqref{adjointAB} we can see that,  for $U=\left(\ba{cc} V_1& 0\\ 0 &V_2\ea\right)\in \rmu(2N)^2$ (the maximal compact subgroup) the ground state is invariant up to an 
irrelevant phase, more precisely $\hat{\mathcal{U}}(U)|\psi_\mathbb{n}\rangle=\det(V_1)^{-\mathbb{n}}|\psi_\mathbb{n}\rangle$ [this is the Fock space counterpart of \eqref{LWinv}]. 
Therefore, equivalence classes of coherent states can be labeled by points $Z\in \mathbb{D}_{4N^4}$ [i.e., the coset $\rmu(2N,2N)/\rmu(2N)^2$], according to the Iwasawa decomposition \eqref{Iwasawa}. Let as 
denote then coherent states simply by $|\psi_\mathbb{n}^U\rangle=|Z\rangle$. 

\begin{prop} The set of coherent states $\{|Z\rangle, Z\in\mathbb{D}_{4N^4}\}$ is not orthogonal, since
\be
\langle Z'|Z\rangle=\frac{\det(\mathbb{1}_{2N}-Z'^\dag Z')^{\mathbb{n}/2}\det(\mathbb{1}_{2N}-Z^\dag Z)^{\mathbb{n}/2}}{\det(\mathbb{1}_{2N}-Z'^\dag Z)^{\mathbb{n}}},\label{overlapcoh}
\ee
but it is overcomplete and resolves the identity
\be
I=\int_{\mathbb{D}_{4N^4}} |Z\rangle \langle Z| d\mu(Z,Z^\dag),\quad d\mu(Z,Z^\dag)=c_\mathbb{n}\det(\mathbb{1}_{2N}-Z^\dag Z)^{-4N}|dZ|,\label{spc2}
\ee
where $|dZ|$ denotes the standard Lebesgue measure on $\mathbb{C}^{4N^2}$ and $c_\mathbb{n}$ is the normalization constant in \eqref{measurecolorcartan}.
\end{prop}
\noindent\textbf{Proof:} Firstly, let us show that 
\be
\langle \psi_\mathbb{n}|\hat{\mathcal{U}}(U)|\psi_\mathbb{n}\rangle=\det(A)^{-\mathbb{n}}, \quad U=\left(\ba{cc} A& C
\\ B &D\ea\right).
\ee
Indeed, let us use the Cartan decomposition $U=VW\tilde V$ with $V=\mathrm{diag}(V_1,V_2)$ and $\tilde V=\mathrm{diag}(\tilde V_1,\tilde V_2)$ block diagonal matrices and 
$V_{1,2}, \tilde V_{1,2}\in \rmu(2N)$. We already know that 
\[\hat{\mathcal{U}}(V)|\psi_\mathbb{n}\rangle=\det(V_1)^{-\mathbb{n}}|\psi_\mathbb{n}\rangle,\; \hat{\mathcal{U}}(\tilde V)|\psi_\mathbb{n}\rangle=
\det(\tilde V_1)^{-\mathbb{n}}|\psi_\mathbb{n}\rangle.\]
Let us factor $W$ as
\[W=\left(\ba{cc} \mathbb{1}_{2N}& 0\\ X &\mathbb{1}_{2N}\ea\right)
\left(\ba{cc} R& 0\\ 0&R^{-1}\ea\right)\left(\ba{cc} \mathbb{1}_{2N}& X\\ 0 &\mathbb{1}_{2N}\ea\right)=U_KU_DU_P.\]
Since the exponential of exciton-annihilation operators  $\hat{\mathcal{U}}(U_K)=\exp[\tr(\hat B X\hat A)]$ contains $\hat B$, its does not affect $|\psi_\mathbb{n}\rangle$. 
In the same manner, the  exponential of exciton-creation operators $\hat{\mathcal{U}}(U_P)=\exp[-\tr(\hat A^\dag X\hat B^\dag)]$ contains $\hat B^\dag$ and it does not affect 
$\langle \psi_\mathbb{n}|$. The only not trivial action comes from  $\hat{\mathcal{U}}(U_D)=\exp[-\tr(\hat A^\dag\ln(R)\hat A)-\tr(\hat B^\dag \ln(R)\hat B)]$. 
Since $\hat{B}^\dag\hat{B}=\hat{B}\hat{B}^\dag-2N\mathbb{1}_{2N}$ and we have the constraint \eqref{qconst}, we arrive to 
$\hat{\mathcal{U}}(U_D)|\psi_\mathbb{n}\rangle=\det(R)^{-\mathbb{n}}|\psi_\mathbb{n}\rangle$. Since $V_1R\tilde V_1=A$, we conclude that
$\langle \psi_\mathbb{n}|\hat{\mathcal{U}}(U)|\psi_\mathbb{n}\rangle=\det(A)^{-\mathbb{n}}$, as desired. Now
\bea 
\langle Z'|Z\rangle&=&\langle \psi_\mathbb{n}|\hat{\mathcal{U}}^\dag(U')\hat{\mathcal{U}}(U)|\psi_\mathbb{n}\rangle=\langle \psi_\mathbb{n}|\hat{\mathcal{U}}(U'^{-1}U)|\psi_\mathbb{n}\rangle=
\det(A'^\dag A-B'^\dag B)^{-\mathbb{n}}\nonumber\\
&=&\det(A'^\dag)^{-\mathbb{n}}\det(\mathbb{1}_{2N}-A'^{-1\dag}B'^\dag BA^{-1})^{-\mathbb{n}}\det(A)^{-\mathbb{n}}.\label{overlapre} 
\eea
Noting $Z=BA^{-1}$ and $Z'=B'A'^{-1}$, as in \eqref{Iwasawa}, and using the phase freedom, we can take $A$ and $A'$ so that  
$\det(A)^{-\mathbb{n}}=\det(AA^\dag)^{-\mathbb{n}/2}=\det(\mathbb{1}_{2N}-Z^\dag Z)^{\mathbb{n}/2}$ (similar for $A'$). In this way, the equation \eqref{overlapre} reduces to 
\eqref{overlapcoh}. 

The closure relation \eqref{spc2} is a consequence of Schur's lemma applied to general unimodular groups, such as $\rmu(2N,2N)$, with square-integrable representations $\blacksquare$

Coherent states defined in \ref{defcoh} also admit a representation in terms of an exponential action of exciton creation operators ${\hat{P}}\equiv \hat A^\dag \hat B^\dag$ on the lowest-weight 
state $|\psi_\mathbb{n}\ra$ as
\be
|Z\ra=\det(\sigma^0-Z^\dag Z)^{\mathbb{n}/2}{e^{\tr({Z}{\hat{P}})}}|\psi_\mathbb{n}\ra.\label{u4cs2}
\ee
This expression resembles the definition of traditional canonical Glauber and Perelomov SU(2) CS as the action of a displacement operator $S(Z)=\det\exp(Z\hat A^\dag \hat B^\dag-\hat B\hat A Z^\dag)$ 
onto the vacuum or lowest-weight state. It also shows the coherent state $|Z\ra$ as a ``Bose-Einstein condensate'' of excitons. 
Interesting physical phenomena of Bose-Einstein condensation of excitons and biexcitons can be found in 
 \cite{excitonBEC}. We believe that our abstract construction of coherent states of excitons can provide 
 an interesting framework to study, not only colored twistors, but also other  possible physical applications in Condensed Matter.
 
The conection with the holomorphic (CS or Bargmann representation) picture presented in section \ref{holopic} is the following. Given an arbitrary state $|\phi\rangle\in \mathcal{H}_\mathbb{n}$ 
of excitons, we define the corresponding CS representation as
\be
\phi(Z)=\langle \phi|Z\rangle \mathfrak{K}_\mathbb{n/2}(Z^\dag,Z),\quad \mathfrak{K}_\mathbb{n/2}(Z^\dag,Z)=\det(\sigma^0-Z^\dag Z)^{-\mathbb{n}/2},
\ee
where $\mathfrak{K}_\mathbb{n/2}(Z^\dag,Z)$ is the square root of the Bergman kernel defined in \eqref{Bergmandef}.  
Indeed, inserting the resolution of the identity \eqref{spc2} in $\langle \phi'|I|\phi\rangle$ we recover the scalar product \eqref{scp2} in 
the Hilbert space $\mathcal{H}_\mathbb{n}(\mathbb{D}_{4N^2})$ of square integrable holomorphic (wave) functions $\phi(Z)$ on the phase space $\mathbb{D}_{4N^2}$. Also, the matrix elements and 
expectation values $\langle Z' |\hat{T}^{\alpha a}|Z\rangle$ (also called \emph{operator symbols}) of colored conformal generators \eqref{bosrepresym} in coherent states $|Z\rangle$ of excitons, reduces 
to simple derivatives of the Bergman kernel $\mathfrak{K}_\mathbb{n}(Z'^\dag,Z)$ in \eqref{Bergmandef}, through the differential realization \eqref{difgen} of colored conformal 
generators on holomorphic functions, as
\be
\langle Z' |\hat{T}^{\alpha a}|Z\rangle=[\mathfrak{K}_\mathbb{n/2}(Z'^\dag,Z')\mathfrak{K}_\mathbb{n/2}(Z^\dag,Z)]^{-1} \mathcal{T}^{\alpha a}\mathfrak{K}_\mathbb{n}(Z'^\dag,Z).
\ee

\section{Field theory}\label{fieldtsec}

Let us now  consider $M$-ctwistors \eqref{Zbf} as matrix fields on Minkowski spacetime $\zeta(x)$. A relativistic invariant filed theory for massive conformal colored particles 
can be proposed by extending the mechanical Lagrangian \eqref{twistL}  to the Lagrangian density 
\be
\tilde\cL=\tr(\partial_\mu\zeta^\dag\Gamma\partial^\mu\zeta),\label{twistL2}
\ee
with the constraint $\zeta^\dag\Gamma\zeta=\mathbb{1}_{2N}$. If we want the $2N$-ctwistors $\zeta(x)$ and $\zeta(x)V(x)$ to be gauge equivalent 
[with local $V(x)\in \rmu(2N)$] transformations) we must perform minimal coupling, replacing  $\partial_\mu\zeta$ by $D_\mu\zeta=\partial_\mu-\ic\zeta \mathcal{A}_\mu$ with 
$\mathcal{A}_\mu$ a $\rmu(2N)$-gauge potential transforming as $\mathcal{A}_\mu\to V^{-1} \mathcal{A}_\mu V-\ic V^{-1}\partial_\mu V$. Therefore, the Lagrangian density 
\be
\cL=\tr[(D_\mu\zeta)^\dag\Gamma D^\mu\zeta],\label{twistLgi21}
\ee
becomes $\rmu(2N)$-gauge invariant. From the equations of motion, we obtain 
$A_\mu=-\ic\zeta^\dag\Gamma\partial_\mu\zeta=\ic\partial_\mu\zeta^\dag\Gamma \zeta$ for the non-dymanical $\rmu(2N)$-gauge vector potential $A_\mu$.  With this 
information, the Lagrangian density \eqref{twistLgi21} 
can be equivalently written as
\be
\cL=\tr\left[\partial_\mu \zeta^\dag \Gamma  P(\zeta,\zeta^\dag)\partial^\mu\zeta\right]=\um \tr \left(\partial_\mu \mathcal{T}\partial^\mu \mathcal{T}\right),\label{twistLgi22}
\ee
with $\mathcal{T}=\zeta\zeta^\dag\Gamma$ a  $\rmu(2N)$-gauge invariant quantity. Expanding $\mathfrak{T}=\mathfrak{T}_a\tilde\lambda^a$, like in \eqref{sunspin}, 
in the Lie algebra basis  $\{\tilde\lambda^a\}$ of the colored conformal group $\rmuu(2N,2N)$, the Lagrangian density \eqref{twistLgi22} can also be written as 
\be
\cL=\frac{1}{2} {\partial_\mu\mathfrak{T}_a}\tilde\delta^{ab}{\partial^\mu\mathfrak{T}_b},\label{sunspinL2}
\ee
which has the form of a nonlinear sigma model (that is, a generalized Heisenberg model of spin-spin
interaction) Lagrangian, with $S_a(x)=\mathfrak{T}_a(x)$ the generalized $\rmu(2N,2N)$ ``spin field'' placed at position $x$, but with a non Euclidean 
metric $\tilde\delta^{ab}$ (see \cite{AOP} for the Euclidean $\rmu(N)$ case and its relation to sigma models for $N$-component fractional
quantum Hall systems). Another equivalent expression for  \eqref{twistLgi22} is given in terms of the minimal fields  $Z\in \mathbb{D}_{4N^2}$ 
though the gauge fixing \eqref{gaugefix} as
\be
\cL= \tr \left(  \Delta_2^2 {\partial_\mu Z} \Delta_1^2  {\partial^\mu Z^\dag}  \right),
\ee
with $\Delta_{1,2}$ in \eqref{cosetcoord}. {In this case, the number of real field degrees of freedom is $8N^2$}.

The field theoretic version of the Berry Lagrangian \eqref{twistLgi3} now adopts the following form 
\be
\mathcal{L}_B=\ic \,\dot{x}^\mu\tr(\zeta^\dag\Gamma\partial_\mu\zeta),\ee
with $\dot{x}^\mu=dx^\mu/d\tau$ and $\tau$ the proper time.  {The number of real field degrees of freedom is now $4N^2$}.

We shall leave the quantization of these ctwistor field theories for future work.

\section{Conclusions and outlook}\label{conclusec}

We have introduced the concepts of colored conformal symmetry $\rmu(2N,2N)$ and colored twistors to address the inclusion of non-Abelian internal $\rmu(N)$ charges into (spinless massive) conformal particles. 
It is known that a genuine $M$-twistor description of a massive particle in four-dimensional Minkowski space fails for the case $M\geq 3$ \cite{nogotwistor}. 
However, replacing standard twistors by colored twistors provides enough room to accommodate non-Abelian internal degrees of freedom, other than the electric charge.

We have analyzed the Lie algebra structure of $\rmu(2N,2N)$ and its discrete series representations, relating the corresponding carrier space to the Hilbert space of a quantized nonlinear 
sigma model of colored free twistors. The quantization is accomplished through a constrained bosonic representation of observables and quantum (Fock) states, which can be defined as particle-hole quanta excitations 
(excitons) above a ground state. We also define coherent states of excitons and the corresponding holomorphic (Bargmann) picture. 

{Whereas the basics and the underlying mathematical (group-theoretical) structure presented here is firm and solid, we recognize that there is still a long way to go to on the physical 
interpretation of phenomenology and final validity of this framework. Before, a reworking of the theory of particle (electromagnetic, weak and strong) interactions, 
rendered in ctwistor terms, including a reframe of gravity, should be presented. This is a long term task, but we think it is worth to explore this proposal. Our next step is to describe 
higher spin colored conformal particles in ctwistor language. The case $N=1$ was treated in \cite{spinning} by extending the Grassmannian $\rmu(2,2)/\rmu(2)\times\rmu(2)$ to the pseudo-flag 
$\rmu(2,2)/\rmu(1)\times\rmu(1)$. There are also other gauged twistor formulations of conformal massive spinning particles in, for example,  \cite{Deguchi1,Deguchi2}. The spinning colored case 
will basically consists in extending $\mathbb{D}_{4N^2}$ to $\rmu(2N,2N)/\rmu(N)\times\rmu(N)$.  }

\section*{Acknowledgments}
This study  has been partially financed by the Consejer\'\i a de Conocimiento, Investigaci\'on y Universidad, Junta de Andaluc\'\i a and European Regional Development Fund (ERDF), Ref. SOMM17/6105/UGR.

\end{document}